\documentclass[aps,twocolumn,showpacs,preprintnumbers,amsmath,amssymb,footinbib,superscriptaddress]{revtex4}

\usepackage{graphicx}

\begin{document}

\title{Linking the chiral and deconfinement phase transitions}

\author{Yoshitaka Hatta} 
\affiliation{Department of Physics, Kyoto University, Kyoto 606-8502, Japan}
\affiliation{The Institute of Physical and Chemical Research (RIKEN)
Wako, Saitama 351-0198, Japan}
\affiliation{High Energy Theory, Department of Physics, Brookhaven National Laboratory, Upton, NY 11973, U.S.A.}

\author{Kenji Fukushima}
\thanks{Present address: Center for Theoretical Physics, Massachusetts
Institute of Technology, Cambridge, MA 02139}
\affiliation{Department of Physics, University of Tokyo, 7-3-1 Hongo, Bunkyo-ku,  Tokyo 113-0033, Japan}

\begin{abstract}
We show that the electric glueball becomes critical at the end-point of the deconfinement phase transition in finite temperature QCD. Based on this observation and existing lattice data, we argue that the chiral phase transition at a zero quark mass and the deconfinement phase transition at an infinite quark mass are continuously connected by the glueball-sigma mixing.
\end{abstract}

\date{\today}

\maketitle

Confinement and chiral symmetry breaking are the
 fundamental properties of quantum chromodynamics (QCD) in the nonperturbative regime.
It has been often speculated how the two phenomena are related to each other.
There are several arguments that the confined vacuum necessarily
 breaks chiral symmetry \cite{casher,thooft}, which suggest that
 the energy scale of chiral symmetry breaking is higher than that of
 confinement. The separation of these scales, i.e.,\
$\sim 4\pi f_{\pi}$ ($f_{\pi}$ is the pion decay constant) for chiral
 symmetry breaking and $\sim \Lambda_{\rm QCD}$ for confinement, has been successfully exploited to reproduce the low energy hadronic properties \cite{Georgi}. 

In view of such experiences at zero temperature, it is quite
surprising that the lattice data at finite temperature suggest that
chiral symmetry restoration and the deconfinement phase transition
occur at the same temperature \cite{Kogut,fukugita,Karsch,aoki,karsch2,gat,allton}. In this
Brief Report we argue how this could be realized.

Fig.~1 is the schematic phase diagram of QCD in the ($T, m$) plane. ($T$ is the temperature and $m$ is the current quark mass.)
The deconfinement phase transition has a definite meaning only at
 an infinite quark mass. In SU(3) pure gluodynamics the
transition is known to be first order \cite{yaffe}. The expectation value
of the Polyakov loop \cite{polyakov,sus},
\begin{eqnarray}
 L(\boldsymbol{x})=\mbox{tr} {\cal P}\exp{\Bigl( -{\rm i}\int^{1/T}_0 A_0(\boldsymbol{x},\tau){\rm d}\tau \Bigr) },\label{11}
\end{eqnarray}
where ${\cal P}$ is the path ordering operator, serves as an order parameter
and jumps at the deconfinement transition temperature. The first order
phase transition persists at a finite but very large quark
mass (solid curve in Fig.~1) and eventually terminates at a second order phase transition
point \textsf{D} where $\langle L(\boldsymbol{x})\rangle$ becomes
continuous.  The universality class of this phase transition is that of the three-dimensional Ising model.
The critical
quark mass at \textsf{D} is estimated to be $\sim{\rm O}(1)$ GeV
in the two-flavor lattice simulations \cite{ha,attig}.

In contrast, chiral symmetry is a true symmetry only at zero quark mass. The order parameter of the chiral phase transition is the quark condensate $\langle \bar{q}q\rangle$.
For three flavors, the phase transition is of first order
\cite{rob}. With finite but small quark masses (which we take to be
equal for all flavors), the first order transition persists and eventually 
terminates at a second order phase transition point \textsf{C}. The jump in the chiral condensate disappears and the sigma meson screening state $\sigma\sim \bar{q}q$ becomes gapless at this point \cite{Gavin,schmidt}.
The location of \textsf{C} and \textsf{D} depends on the number of flavors and their mass ratios. [For example, \textsf{C} lies on the $m=0$ axis for two flavors \cite{rob}.] The following argument does not depend on the details of the flavor sector, however.  

\begin{figure}
\includegraphics{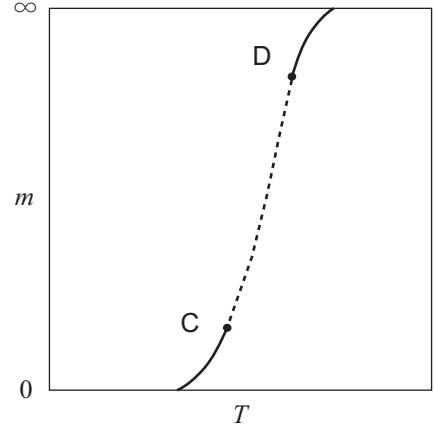}
  \caption{\label{fig} The phase diagram in the ($T$,$m$) plane}
\end{figure}

Between the points $\textsf{C}$ and $\textsf{D}$, there is no true
phase transition. There are just two crossover phase transitions starting
from $\textsf{C}$ and $\textsf{D}$. The location of the crossover transition (the pseudo-critical temperature) is determined 
from the peak position of the susceptibility of each order parameter. The lattice data suggest that the two pseudo-critical temperatures coincide for all values of the quark mass, which means that two crossover
curves are smoothly connected at some intermediate point (dotted curve in Fig.~1).
It is as if the deconfinement phase transition in the pure gluonic theory is `guided' towards the chiral phase transition traveling over an infinite range of the quark mass. 

The nature of this `coincidence' has been discussed in \cite{gock,chen,satz,fuku}, where the authors attributed the (almost) singular behavior of the Polyakov loop to the coupling of the Polyakov loop and the
chiral condensate. It turns out that their observation can be incorporated in a very simple and intuitive picture which enables us to grasp what is going on near the finite temperature QCD phase transition for all values of the quark mass.

Since the center symmetry is explicitly broken by dynamical quarks \cite{quark}, the Polyakov loop has a nonzero expectation value at \textsf{D}.
Thus the phase transition at \textsf{D} would be characterized by the infinite correlation length (or the vanishing screening mass) of the shifted order parameter
\begin{eqnarray}
L'(\boldsymbol{x}) \equiv L(\boldsymbol{x})-\langle L \rangle, \label{l}
\end{eqnarray}
where $\langle L \rangle$ is the expectation value of $L$ at \textsf{D}.
However, there is an apparent puzzle. What is the physical meaning of the `mass' of the $L'$ field? A naive guess is the string tension as in the case of the pure SU(2) theory. This is wrong because the string eventually breaks at large distances in the presence of dynamical quarks no matter how heavy they are, whereas the second order phase transition requires an infinite correlation length. We will show that the massless mode at \textsf{D} is actually the scalar {\it glueball} screening state. 
  
For this purpose, first we work in the pure SU(2) gauge theory. In this case the deconfinement phase transition is second order
in the universality class  of the three dimensional
Ising model. The string tension $K$ goes to zero smoothly as the critical temperature is approached
\begin{eqnarray}
K \sim |t|^{\nu}, \qquad (t<0) \label{sigma}
\end{eqnarray}
where $t$ is the reduced temperature and the universal exponent is $\nu \simeq 0.63$ \cite{Engels}. 

One can realize the relevance of glueballs by considering the specific heat $C_{\rm v}$. In the theory of critical phenomena, the singular part of the specific heat is represented by the integrated correlator of the order parameter squared \cite{zinn},
\begin{eqnarray}
 C_{\rm v} \sim \int {\rm d}^d x \langle
 L^2(\boldsymbol{x})L^2(\boldsymbol{0})\rangle_{\rm c}  \sim |t|^{-\alpha}, \label{c}
\end{eqnarray}
where $\alpha\simeq 0.11$ and $d=3$ is the spatial dimension. The subscript c denotes the connected part. (Note that $L(\boldsymbol{x})$ is real for SU(2) and $\langle L^2(\boldsymbol{x}) \rangle$ stays nonzero at any temperature.)
The scaling dimension of the order parameter squared is $d-1/\nu$ \cite{zinn}. Thus the explicit structure of the divergence is 
\begin{eqnarray}
 \int {\rm d}^d x \frac{{\rm e}^{-m_{\rm G}|\boldsymbol{x}|}}{|\boldsymbol{x}|^{2(d-\frac{1}{\nu})}}\sim |t|^{-\alpha}, \label{c2}
\end{eqnarray}
where we have assumed that the connected part of the correlator in
(\ref{c}) exponentially decays as $\sim {\rm e}^{-m_{\rm G}|\boldsymbol{x}|}$
at large distances.
Matching the powers of $|t|$ and using Josephson's law $\alpha=2-d\nu$ (hyperscaling), we obtain,
\begin{eqnarray}
m_{\rm G}\sim |t|^{\nu}. \label{ttt}
\end{eqnarray}
Hence the operator $L^2$ couples to a state with zero screening mass at the phase transition. 
By using the SU(2) identity 
\begin{eqnarray}
L^2(\boldsymbol{x})=1+L_{\rm A}(\boldsymbol{x}),
\end{eqnarray}
where $L_{\rm A}$ is the Polyakov loop in the adjoint representation (i.e., the trace in Eq.~(\ref{11}) is taken in the adjoint representation), or just by 
noticing that $L^2$ is invariant under the center transformation, $L
\to -L$, we see that the massless state is the glueball 
\cite{glue}. [More precisely, the {\it electric} glueball screening state. See below.]

It should be noted that Eq.~(\ref{ttt}) does {\it not} mean that the (electric) glueball is another order parameter field. Rather, it is a trivial consequence of the scaling property at general second order phase transitions; the only relevant scale is the correlation length $\xi \sim |t|^{-\nu}$. [The word 'massless' may be somewhat misleading since we normally associate it with the order parameter field, namely, the Polyakov loop.]

One might suspect that the vanishing of the glueball screening mass would contradict with the presence of the trace anomaly. In the simple phenomenological Lagrangian of glueballs at zero and low temperatures \cite{sch,ellis}, the glueballs acquire their mass entirely from the trace anomaly, or the gluon condensate. However, the exact lattice sum rule for the glueball mass \cite{rothe} shows that only $\frac{1}{4}$ of the glueball mass comes from the expectation value of the trace anomaly in the glueball state measured {\it relative} to the vacuum. Moreover, the scalar glueball mass at zero temperature splits into two branches at finite temperature.  
The glueballs which become massless at the phase transition are the  electric glueballs (those containing $A_0$, or time-like links), while the screening mass of the magnetic glueball (composed of $A_i$, or space-like links) remains finite. This crucial difference has to be taken into account in phenomenological models such as \cite{sannino}. 
As long as the specific heat diverges in the usual manner at the SU(2) deconfinement phase transition \cite{cv}, we believe that the vanishing of the electric glueball screening mass and the non-vanishing of the trace anomaly are compatible.  A significant decrease of the electric glueball mass near the
SU(2) deconfinement transition has been observed on the lattice \cite{datta}.

Let us go back to the end-point \textsf{D} of SU(3) QCD with dynamical quarks. We can think of an effective Ginzburg-Landau theory near \textsf{D} in terms of the new order parameter $L'$ defined in Eq.(\ref{l}).
However, due to our identification
\begin{eqnarray}
G(\boldsymbol{x}) \sim |L(\boldsymbol{x})|^2=|L'(\boldsymbol{x})+\langle L \rangle |^2,
\end{eqnarray}
the glueball field $G(\boldsymbol{x})$ contains a linear term in $L'$. Hence the screening mass of the Polyakov loop which vanishes at \textsf{D} is equal to the screening mass of the glueball, and we may as well construct a Ginzburg-Landau theory of $G$ around \textsf{D}. [Note that in contrast to the SU(2) case, the electric glueball is {\it the} order parameter field in the case of SU(3).]  The crucial point is that, {\it in the presence of dynamical quarks, the electric glueballs must mix with the $\bar{q}q$ states}. In fact, the correct massless field associated with the second order phase transition at \textsf{D} is a linear combination of the
pure glueball interpolating field $G$ and the sigma meson interpolating field $\sigma$ with large gluonic content,
\begin{eqnarray}
\phi=G\cos \theta  + \sigma\sin \theta, \ \ \ \ \ \sin \theta \ll 1.\label{g}
\end{eqnarray}
 The other linear
combination with large mesonic content,
\begin{eqnarray}
\phi'=-G\sin \theta + \sigma \cos \theta,
\end{eqnarray}
is massive, hence decouples at \textsf{D}.

The mixing Eq.~(\ref{g}) inevitably occurs if the first order phase transition line makes an angle with the quark mass axis at \textsf{D}, and there is no symmetry to keep them parallel. In the same way, at \textsf{C} the critical sigma screening state slightly mixes with the glueball screening state. The simultaneous enhancement of the chiral susceptibility
\begin{eqnarray}
\chi_{\rm ch} \sim  \int {\rm d}^d x\langle \bar{q}q(\boldsymbol{x})\bar{q}q(\boldsymbol{0})\rangle,
\end{eqnarray}
and the Polyakov loop susceptibility along {\it each} crossover phase transition line is a trivial consequence of the mixing \cite{satz,fuku}.

Now we address a deeper question: The lattice data suggest that there is only {\it one} crossover phase transition line in the $(T, m)$ plane, not two.
With the above line of argument we are led to the following scenario which naturally explains this longstanding puzzle.
Let us decrease the quark mass $m$ and follow the crossover phase
transition line starting from \textsf{D}. The screening
mass of the $\phi$ field first increases from zero with a certain
critical exponent. At the same time, the mixing angle $\theta$
gradually increases. If the mixing angle stops growing and stays small, the $\phi'$ field, which is then always sigma-like, serves as the order parameter at \textsf{C}. In this case, the two phase transitions would have nothing to do with each other and there would be two peaks in {\it both} the chiral and Polyakov loop susceptibilities. [Remember that the coincidence occurs irrespective of whether or not the two crossover lines meet.]
However, the observed single peak means that {\it the $\phi$ field describes the field with the smallest screening mass (or
the longest correlation length) in the system for all values of the quark mass and
the mixing angle gradually changes from $\theta \sim 0$ to $\theta
\sim \pi/2$.} 
Namely, the $\phi$ ($\phi'$) field is
first glueball-like (sigma-like) for large values of the quark mass and it becomes sigma-like (glueball-like) as the quark mass gets
smaller. In order for this to be realized, a {\it level repulsion} between the $\phi$ state and the $\phi'$ state must occur at an intermediate value of the quark mass.
 The fate of the $\phi$ field is predominantly sigma-like
and is nothing but the order parameter of the genuine QCD phase
transition, the chiral phase transition. This gives a simple and intuitive explanation of why there seems to be only one phase transition in finite temperature QCD lattice simulations.

An immediate consequence of the mixing Eq.~(\ref{g}) is that the electric glueball ({\it or} the Polyakov loop) and the sigma meson screening masses {\it coincide} at the common pseudo-critical temperature of the crossover phase transition. [Note that in principle, the mixing between the sigma and the glueball could occur in the entire phase diagram (Fig.~1). However, we expect that in most regions the mixing is too small to be observed on the lattice. For example, above the chiral phase transition temperature, at small quark masses the sigma meson and the pions fall in a multiplet due to (approximate) chiral symmetry. The glueball screening mass is indeed different from the sigma screening mass in this region \cite{gavai}. Also, at low temperatures the glueball will primarily mix with nearby scalar quarkonium states. The point is that near the crossover phase transition, the sigma meson and the glueball conspire to create the lightest particle which drives the phase transition. At large distances, no other states can couple to this mode. Thus the glueball-sigma mixing and the resulting degeneracy in their  screening mass should be observable in this region (but not necessarily in other regions).]
Detailed lattice measurements of the correlators (or the off-diagonal correlators) of electric glueballs, Polyakov loops and sigma mesons around the region between \textsf{C} and \textsf{D} would give us the information of the lowest and the second lowest screening masses as well as the quark-mass dependence of the  mixing angle. Compilation of such measurements would confirm or exclude our scenario.

It is intriguing to consider the phase diagram for different values of $N_{\rm c}$, the number of colors.
As $N_{\rm c}$ increases, it is likely that the first order deconfinement phase transition strengthens \cite{teper} and the point \textsf{D} moves downward. Presuming that our scenario also holds for general $N_{\rm c} \ge 3$, we expect that the point \textsf{D} {\it reaches} the point \textsf{C} at some value of $N_{\rm c}=N_{\rm c}^*$ and the high and low temperature phases are separated by a single first order phase transition line for $N_{\rm c}\ge N_{\rm c}^*$. This is consistent with the early observations \cite{neri}.
From our viewpoint, $N_{\rm c}=2$ is special. The Ginzburg-Landau theory of $G$ does not make sense at \textsf{D} which now lies on the $m=\infty$ axis. Also the nature of the chiral phase transition is peculiar for $N_{\rm c}=2$ due to the Pauli-G\"{u}rsey symmetry \cite{nishida}. It is thus interesting to pursue a similar link (glueball-sigma-diquark mixing) between the two phase transitions in $N_{\rm c}=2$ QCD at finite temperature and baryon density.

In conclusion,  based on the existing lattice data, we have proposed a novel scenario of the finite temperature QCD phase transition. The chiral phase transition at zero quark mass and the deconfinement phase transition at infinite quark mass are continuously connected by the glueball-sigma mixing which is a well-known phenomenon in the zero temperature scalar meson spectrum. In other words, the two phase transitions are just different limits of a single (crossover) phase transition driven by a single field.
Phenomenological consequences of this scenario remain to be explored.

{\bf Acknowledgements:}

We thank M.~Creutz, L.~McLerran, P.~Petreczky, R.~D.~Pisarski, K.~Splittorff and M.~A.~Stephanov for their interest in this work and helpful comments.
Y.~H.\ thanks RIKEN BNL Reserch Center. K.~F.\ is supported by the Japan Society for the Promotion of Science for Young Scientists.


\begin{thebibliography}{999}
\bibitem{glue}
The state created by the operator $L_{\rm A}$ is often referred as a `gluelump' in distinction from a `glueball' \cite{michael}. However, we expect that the difference vanishes near the critical point. Note that in practical lattice simulations the specific heat is measured from the integrated correlator of plaquettes which excite glueballs.
\bibitem{cv}
Due to the very small critical exponent, $\alpha \sim 0.11$, the correct singularity in the specific heat has not been confirmed on the lattice so far. See, however, \cite{eng}.
\end{thebibliography}
\end{document}